\documentclass{article}
\usepackage{arxiv}

\usepackage[utf8]{inputenc} 
\usepackage[T1]{fontenc}    
\usepackage{hyperref}       
\usepackage{url}            
\usepackage{booktabs}       
\usepackage{amsfonts}       
\usepackage{nicefrac}       
\usepackage{microtype}      
\usepackage{lipsum}		    
\usepackage{graphicx}
\usepackage[square,sort,comma,numbers]{natbib}
\usepackage{doi}
\usepackage{multirow}
\usepackage{balance}
\usepackage{threeparttable}
\usepackage{subcaption} 
\usepackage{kotex}

\title{An Empirical Analysis on Transparent Algorithmic Exploration in Recommender Systems}
\author{Kihwan Kim\\
Department of Computer Science and Engineering\\
UNIST\\
\texttt{kh1875@unist.ac.kr} \\
}
\renewcommand{\headeright}{}
\renewcommand{\undertitle}{}

\begin{document}
\maketitle
\begin{abstract} 
All learning algorithms for recommendations face inevitable and critical trade-off between exploiting partial knowledge of a user's preferences for short-term satisfaction and exploring additional user preferences for long-term coverage.
Although exploration is indispensable for long success of a recommender system, the exploration has been considered as the risk to decrease user satisfaction.
The reason for the risk is that items chosen for exploration frequently mismatch with the user's interests. 
To mitigate this risk, recommender systems have mixed items chosen for exploration into a recommendation list, disguising the items as recommendations to elicit feedback on the items to discover the user's additional tastes.
This mix-in approach has been widely used in many recommenders, but there is rare research, evaluating the effectiveness of the mix-in approach or proposing a new approach for eliciting user feedback without deceiving users. 
In this work, we aim to propose a new approach for feedback elicitation without any deception and compare our approach to the conventional mix-in approach for evaluation. 
To this end, we designed a recommender interface that reveals which items are for exploration and conducted a within-subject study with 94 MTurk workers.  
Our results indicated that users left significantly more feedback on items chosen for exploration with our interface. 
Besides, users evaluated that our new interface is better than the conventional mix-in interface in terms of novelty, diversity, transparency, trust, and satisfaction.
Finally, path analysis show that, in only our new interface, exploration caused to increase user-centric evaluation metrics.
Our work paves the way for how to design an interface, which utilizes learning algorithm based on users' feedback signals, giving better user experience and gathering more feedback data.
\end{abstract}
\section{Introduction}
Recommender systems have helped users navigate numerous items~\cite{Resnick97} (e.g., news content, commercial products, and movies) in many web sites. 
In particular, commercial web sites extensively utilize recommender systems to persuade customers to buy items and return for future purchases~\cite{Haubl00}. 
To improve the effectiveness of recommender systems, the exploitation approach has been used~\cite{Ricci15}, aiming at producing recommendations that are best matched to user preferences. 
In this approach, the usefulness of recommender systems is evaluated by measuring accuracy of recommendations on user preference (e.g., how many items shown to users are matched to users' preference?). 
While effective, this approach often results in reduced diversity in recommendations~\cite{Vargas14, Wu16, Kaminskas17, Kotkov18}), as most items are similar to each other. 
Such weakness on diversity can make users lose interest on recommendations, reducing not only the number of user feedback for extracting hidden user preference, but also perceived usefulness and satisfaction of recommendations~\cite{Pu11}.

To complement the weakness of the exploitation approach, the algorithmic exploration approach is utilized~\cite{Chen16}.
Originating from \textit{online learning}~\cite{Shalev12}, the exploration approach allows a recommender system to try out new items (i.e., `exploratory items'). 
The rationale behind presenting exploratory items is that users often interact with the items, and the logs produced by the interactions can be used for improving the coverage of the user preference information and performance of recommender systems~\cite{Bottou13,Li11,Swaminathan15}.
By utilizing the algorithmic exploration, recommender systems can enhance the quality of recommendations with respect to not only a single user, but also other users, for example, through exploring newly released items~\cite{Aharon15,Anava15}.

Although the exploration approach can be useful for improving recommender systems, it also has potential to decrease user satisfaction, because exploratory items may not reflect user preferences.
To resolve this issue, recommender systems utilize mix-in lists, where exploratory items disguise as recommendations~\cite{Schnabel18, Zheng18, Ma16} and are mixed into genuine recommendation lists.
As the mix-in list becomes popular in the recommender systems~\cite{Schnabel18, Zheng18, Aharon15, Wang14, Li10}, it is important to understand how effective the mix-in list in terms of user satisfaction.

Measuring user satisfaction of recommender system is not straightforward, as user satisfaction does not always correlate with high accuracy of recommendation algorithms. 
Thus much research has been performed with aims for improving recommender systems of using mix-in lists from users' point of view~\cite{Ge10,Mcnee06,Kaminskas17,Pu11}.
Prior research investigates qualitative aspects of recommender systems, including systems' usability, usefulness, interaction and interfaces, and user satisfaction and finds out if any of the aspects impact on users' behaviors. 
A study that is directly related to the evaluation is Schnabel et al.'s work~\cite{Schnabel18} that shows users can perceive the list is not best tuned for their preference, according to the number of exploratory items. 
They also show that the mix-in list can decrease trust and transparency of the recommender systems.
However, due to limited number of studies for the qualitative evaluation on recommended items, we do not have clear understanding of the mix-in list and several questions on the mix-in approach remain unanswered.
For example, there are not answers on 1) whether deceiving users brings the best result for recommender systems, 2) how to resolve the issue of possible worsened user satisfaction with the deception, and 3) how to enhance user experience with exploratory items.

\begin{figure}[t]
  \centering
  \includegraphics[width=0.95\textwidth]{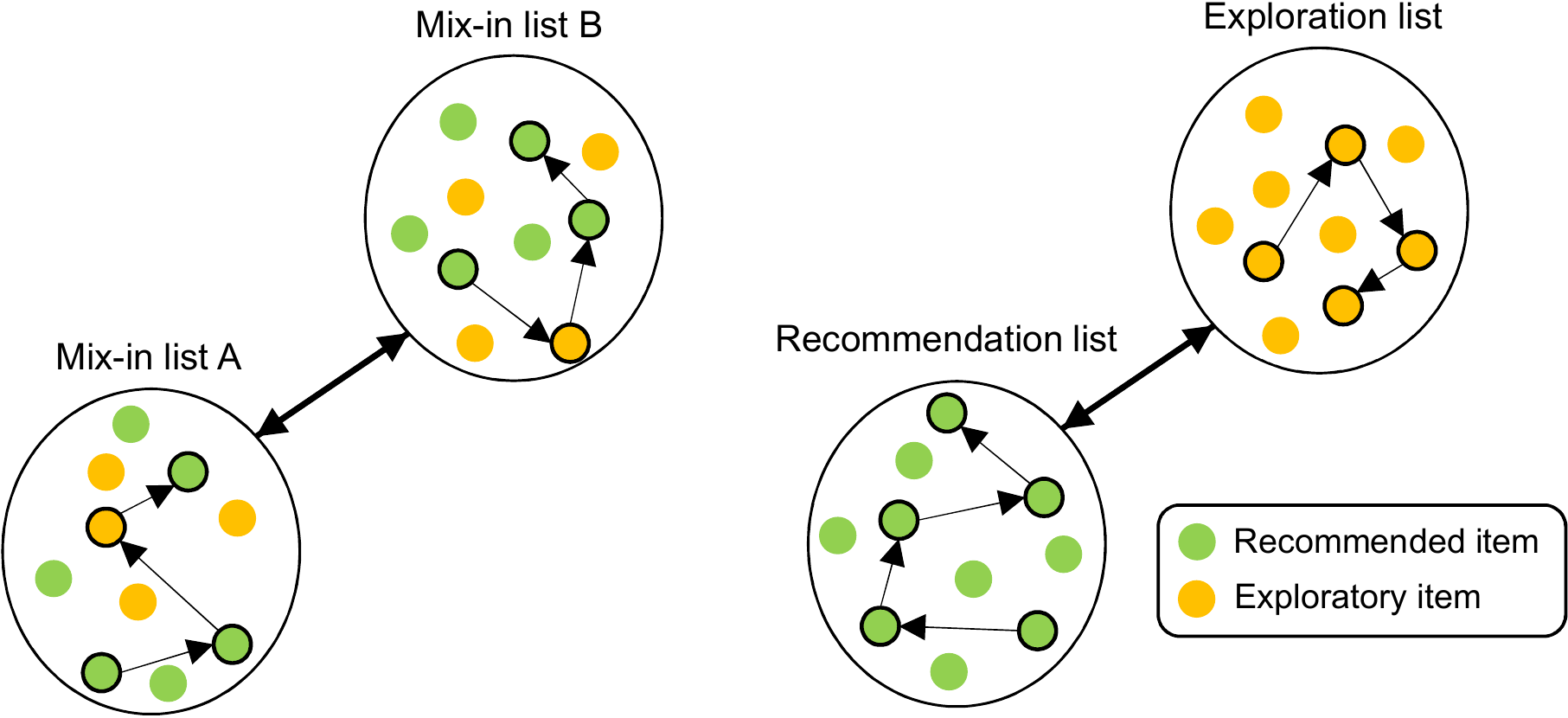}
  \caption{Two interfaces are used in our study. Mix-in interface (left) blends exploratory items with recommended items in the recommendation lists. Revealed interface (right) informs users that the exploration list has exploratory items only.}
  \label{fig_exp}
\end{figure}

In this work we aim to answer the three questions on the mix-in list and find a method to improve effectiveness of recommender systems in terms of user-centric objectives without degrading accuracy. 
To achieve the goal, we design a novel recommendation interface, called a revealed interface, where recommended and exploratory items are provided in two separated lists, in order to let users know which items are exploratory ones during their navigation. 
\autoref{fig_exp} (right) shows a navigation example, where users explore the items recommended first for finding items for  preference and then jump to another list for reviewing exploratory items.  

To evaluate the impact of the proposed interface, we conducted a within-subject experiment with 94 online participants recruited from Amazon Mechanical Turk (MTurk).
In the experiment, we asked to the workers to choose a movie to watch with an interface between \textbf{mix-in} and \textbf{revealed} interfaces at a time. 
With the revealed interface users could access recommended items and exploratory items in two separated lists--mix-in and revealed lists, informing users which list has exploratory items. 
In contrast, when users used the mix-in interface, they were allowed to navigate two conventional mix-in lists, where exploratory items are inserted into the recommendation lists. 
The experimental results indicate that the newly proposed revealed interface received higher scores than conventional mix-in interface in terms of diversity, novelty, transparency, trust, and user satisfaction of a recommender system.  
We also find that the newly proposed revealed interface collected more interaction logs on exploratory items without impacting accuracy of a recommender system than the conventional mix-in lists. 
Lastly, we conduct a multi-group path analysis and find out that the proposed interface improved 1) novelty, diversity, and transparency and 2) the improved diversity enhances user satisfaction of a recommender system without impacting accuracy. 

The contributions of this work include that the design of a novel recommender system interface uses an opposite recommendation presentation strategy, an experiment for analyzing the effect of the mix-in and revealed interfaces, and a multi-group path analysis for causal relationship analysis among user-centric objectives.

The remainder of this paper is structured as follows: 
In the next section (\autoref{sec_relatedwork}), we describe prior work on algorithmic exploration, user-centric objectives for recommender systems, and strategies for recommendation presentation. 
Then we present our research questions in \autoref{sec_rq}, followed by our experiment design (\autoref{sec_study}). 
In Section~\ref{sec_result}, we analyze the experiment data in user-centric objectives and user feedback quantity perspectives. 
Then we give implications of this work (\autoref{sec_discussion}) and discuss the limitations and future work of this work (\autoref{sec_limitations}).
Finally, we conclude this work (\autoref{sec_conclusions}). 

\section{Related Work} 
\label{sec_relatedwork}
In this section we review prior work on algorithm exploration, user-centric objectives, and recommendation presentation strategies. 

\subsection{Collecting User Feedback and Algorithmic Exploration}
As recommender systems aim at providing recommendations that users would most like, collecting user preference information is crucial.
Much research exists on what types of user feedback could be useful for understanding user preference and producing accurate recommendations, in addition to conventional user feedback from mouse clicks~\cite{Radlinski08}.
Examples include the studies on measuring the impact of using gaze~\cite{Huang11}, dwell time~\cite{Kim14, Liu10}, cursor movements~\cite{Huang12,Huang11,Zen16}, and scrolling~\cite{Guo12,Liu17} as user feedback signals.
A hybrid approach is also proposed combining collaborative filtering and content-based recommendations~\cite{Cohen17}.
Specifically, this approach maps the feature space of items to collaborative filtering's latent space to approximate new items' latent vectors~\cite{Agarwal09,Gantner10} and can reduce the impact of the cold start problem. 

Algorithmic exploration is a concept in \textit{online learning~}~\cite{Shalev12}, where algorithms learn sequentially, and try out new actions or options to widen the learning domain coverage in the long term.
As the approach extracts additional information on users, recommender systems actively adopt ~\cite{Wang14, Zheng18, Aharon15, Anava15} the algorithmic exploration for discovering unrevealed user preference and increasing catalog coverage.
While effective, it is considered that algorithmic exploration may decrease user experience, since the items chosen for exploration frequently turn out to mismatch the user’s interests~\cite{Schnabel18}.
In other words, it is possible that algorithmic exploration leads to user dissatisfaction due to low accuracy in recommendations.
To make algorithmic exploration be more successful, Aharon~\cite{Aharon15} propose smart exploration by gradually excavating users who  are more likely to be interested in the new items and matching the new items based on users’ interactions.
While previous work focuses on how to choose items or users, we aim at designing a novel interface for presenting algorithmic exploration items so that more feedback can be gathered. 
This implies that our interface can be used with existing methods for collecting more user feedback.

\subsection{User-centric Objectives for Recommender Systems}
Evaluating recommender systems with accuracy~\cite{Adomavicius05,Herlocker04} alone is not sufficient to fully describe how users perceive recommendations~\cite{Ge10,Mcnee06,Kaminskas17,Pu11}. 
Thus, recent research starts investigating other objectives beyond accuracy that can capture qualitative aspects of recommender systems. 
The newly proposed objectives include novelty, diversity, transparency, and satisfaction.
 
Here, novelty and diversity are used for measuring in a user perspective how unknown recommended items are~\cite{Mcnee06} and for gauging how dissimilar recommended items are~\cite{Vargas14}, respectively.
Transparency~\cite{Sinha02} determines whether or not a recommender system allows users to understand why a particular item is recommended to them, while trust~\cite{Donovan05} indicates whether or not users find the whole system trustworthy. 
Next we describe prior studies on the novelty and diversity.

Existing work has demonstrated that improving novelty and diversity of recommendations lead to better user experience with very little harm to perceived accuracy.
As there is a trade-off relationship between algorithmic accuracy and novelty/diversity of recommendations, researchers argue that recommender systems should manage how to coordinate them~\cite{Zhou10, Javari15}. 

A basic strategy for such coordination is to compute similarity scores between users and items and set a threshold similarity score for selection.
Then items are randomly picked that have a similarity score higher than the threshold. 
With this strategy, when the threshold is high, the recommendations seem identical to general similarity-based computation results. 
In contrast, if the threshold is set to a very low score, the presented recommendations can be viewed as a group of randomly selected items.
This observation implies a trade-off relationship in that if we use higher thresholds, we cannot expect improved novelty and diversity and with low threshold scores, we have decreased recommendation accuracy~\cite{Pu11}.
As such much research has been performed on the relationship and to find appropriate thresholds for achieving high novelty and diversity without impacting accuracy~\cite{Smyth01, Ziegler05, McSherry02, Hurley11}. 
For example, Smyth and McClave~\cite{Smyth01} propose three heuristic algorithms that combine similarity and diversity--bounded random selection, greed selection, and bounded greed selection. 
Their experiments report that the greed selection method shows the best result that adds items at a time to the recommendation list according to the computed score.
Ziegler et al.~\cite{Ziegler05} also showcase a similarity computation algorithm that compute an intralist similarity score based on overall diversity of the recommendation list. 

The algorithmic approaches for improving novelty and diversity of recommender systems are fundamentally the same in that they randomly pick items with the scores higher than the threshold to guarantee intended accuracy. 
But what is different between the two objectives are the perspectives that they focus on--novelty is decided by how much users do not know the received items and diversity is determined by how different the items are from each other in the list. 

Several studies exist on exploring relationships among the objectives. 
Schnabel et al.~\cite{Schnabel18} perform experiments and demonstrate that transparency negatively correlates with the amount of exploration, while novelty does positively. 
In this work, the exploration process is hidden to users and the condition of the experiment is only the amount of exploration. 
We also analyze the relationship, but the experiment condition is different--users know the exploration process. 
As our approach is to let users know the exploration process, we think that transparency~\cite{Sinha02} and trust~\cite{Donovan05} are the most relevant objectives to our work. 
Pu et al.~\cite{Pu11} show that the two objectives are closely related; transparency significantly impacts trust. 
Recent research provides visual interfaces for eliciting preference, such as slider bars and finds that the interfaces can make users perceive controllability on the systems and more satisfied~\cite{Bostandjiev12,Chang15,Harper15}. 
Motivated by the idea that interfaces can affect how users perceive recommender systems, we design a new interface, conduct experiments, and analyze relationship between the objectives in this work. 

\subsection{Strategies for Recommendation Presentation}
As users interact with recommender systems, properties of recommendation sets (e.g., item presentation strategies and the number of random items in recommendation lists), could affect effectiveness of recommender systems in several perspectives, such as perceived accuracy, diversity, user satisfaction, and purchase intention.
For example, Ge et al.~\cite{Ge12} show that placing diverse items at the end of recommendation lists increase users' perceived diversity of recommendation lists. 
Schnabel et al.'s experiment~\cite{Schnabel18} is another example that discloses that the number of exploration items in the recommendation lists affects user perception of interactive session-based recommender systems and the quantity of implicit feedback.
The experiment also reveals that when a recommendation list provides many exploration items (e.g., more than 60\% of the total items~\cite{Schnabel18}), it significantly undermines users' perceived quality on the recommendations.
This result can be understood that letting users know which items for exploration does not give any benefit in practice, although other studies that show transparency plays a critical role in user experience of recommender systems~\cite{Amershi19,Sinha02,Pu11}.
Compared to the previous work, we systematically design an interface for transparency, perform an experiment, and find that transparently revealing exploration items to users is a better presentation strategy than the conventional approach of hiding exploration from users in terms of user experience.  

\section{Research Questions}
\label{sec_rq}
Conventionally, exploratory items are not chosen based on user profile or preference, but randomly selected.
This often results in mismatches between recommendations and user interests. 
Thus, many recommender systems have assumed that exploratory items are risky and undermine user experience. 
But when recommender systems exclude exploratory items, they end up having a narrow coverage in recommendations, as limited information on user preference always leads to same recommendations.
To resolve the dilemma, recommender systems have started using mix-in recommendation lists, where exploratory items are mixed into an existing recommendation list. 
By using the mix-in list, recommender systems expect to not only attract users with recommended items, but also collect additional user feedback or hidden user interests with exploratory items.
Many studies show the importance of the information collected from the exploratory items in the mix-in list in widening their recommendation coverage~\cite{Bottou13,Li11,Swaminathan15}.

This work starts from a doubt whether using the mix-in list can be considered as an optimal solution in terms of collecting user feedback.
In particular, we hypothesize that there could be other ways to collect more user feedback with changes in interfaces. 
Thus our first question is \textbf{how to design an interface for recommender systems that allows us to collect more user feedback than conventional mix-in interfaces without impacting accuracy of recommendations (Q1)}. 
We introduce our interface design in \autoref{sec_interface} that informs users of exploratory items in recommendation lists. 

As our interface is new, the next question is \textbf{how users perceive the proposed interface (Q2)}. 
This means, if users think the interface is worse than the existing mix-in interface in accessing recommendations, the interface cannot be worth, even if it enables more user feedback. 
To evaluate the interface, we consider to use the conventional metric, accuracy with additional evaluation perspectives (i.e., beyond-accuracy objectives), such as transparency, diversity, novelty, trust, and satisfaction. 
As our interface is designed to reveal exploratory items to users, we assume that it increases systems' transparency and can help users understand why the items are presented to them, allowing users to feel the system is trustful~\cite{Schnabel18, Barraza17, Pu11} and accurate~\cite{Schnabel18, Barraza17}.
Lastly, we expect a more number of user interactions on the exploratory items than the conventional mix-in interface due to two possible cases--users who want items out of their usual preference and those who know most of the items in the recommendations. 

\begin{figure*}[t]
  \centering
  \includegraphics[width=0.95\textwidth]{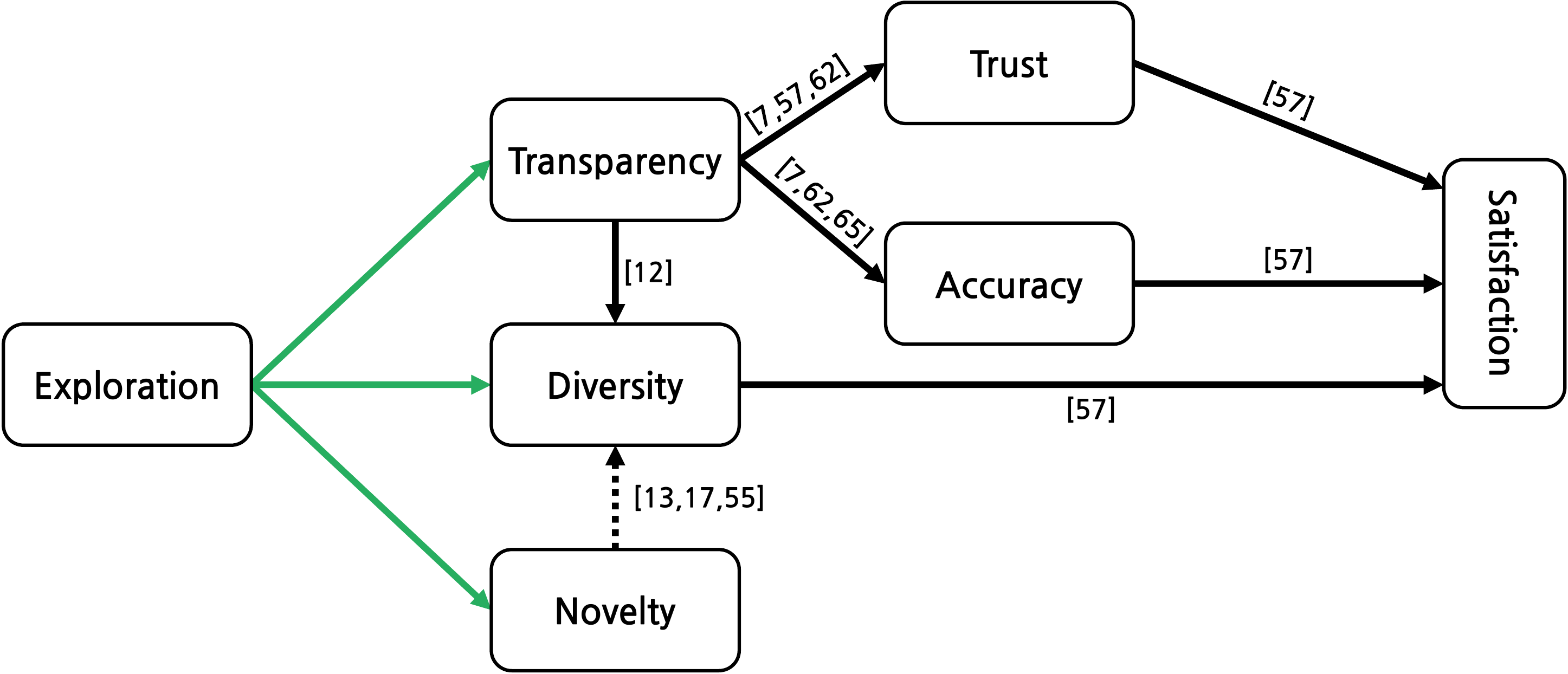}
  \caption{A summary of previous work on user-centric objectives. The black paths with arrows mean relationship between two objectives found from literature. The dotted lines represent a controversial relationship. We aim to investigate the relationship between objectives shown with green paths.}
  \label{proposed_path_model}
\end{figure*}

Several studies explore the relationships between the user-centric objectives for recommender systems. 
A summary of the previous work on the objectives is presented in \autoref{proposed_path_model}.
At first, we see that Pu et al.~\cite{Pu11} report that improving diversity, trust, and accuracy of recommender system positively influence user satisfaction. 

Simonson~\cite{Simonson05} argues that transparency plays an important role in improving perceived accuracy recommender systems. 
This means, when users see a correspondence between the preferences expressed in the measurement process and the recommendations presented by the system, they would consider given recommendations are accurate. 
Two reports exist on diversity. 

Castagnos et al.~\cite{Castagnos13} argue that helping users understand why a particular item is recommended to them allows them to perceive diversified recommendations (i.e., increasing transparency improves diversity).
The relationship between novelty and diversity seem somewhat conditional. 
First, Ekstrand et al.~\cite{Ekstrand14} propose that improving novelty positively impacts diversity.
But Castells et al.~\cite{Castells11} argue that such positive relationship between the two objectives can emerge in a long-term use of recommender systems. 
This implies that we may not observe the relationship in session-based recommender systems. 
Niu et al.~\cite{Niu18} also claim that we need to be careful when investigating novelty and diversity--all diverse items are not novel and all novel items are not diverse. 
The relationship between novelty and satisfaction is also controversial.
For example, we find that Ekstrand et al.~\cite{Ekstrand14} argue that novelty negatively impacts on satisfaction, while other studies report positive impact of novelty to satisfaction~\cite{Chen19, Pu11}. 
But the researchers agree that novelty has an indirect positive effect on satisfaction through diversity~\cite{Ekstrand14,Chen19, Pu11}. 
Thus we reflect the indirect impact of novelty to satisfaction in \autoref{proposed_path_model}. 

Pu et al.~\cite{Pu11} perform a survey called ResQue and reports that all of the diversity, trust, and diversity are positively related to satisfaction. 
In the survey, the participants need to answer the survey questions by remembering their prior experience of different online recommender systems. 
As such there is a need to perform an experiment with a single recommender system to confirm the relationship among the diversity, trust, accuracy and satisfaction. 

Given the existing results on the relationships among user-centric objectives, our last question is on \textbf{the impact of the proposed interface on the user-centric objectives (Q3)}.
Specifically, we are interested in investigating the strength and significance of the relationships among objectives in the two interfaces.
For example, we aim to find whether the mix-in approach decreases transparency, and if so, we can estimate other consecutive impacts on the objectives by following the relationship paths. 
We can also question that if users perform more user interactions with exploration items in the proposed interface as intended, what impact we can observe in terms of diversity and novelty objectives.

We summarize our research questions as follows:
\begin{itemize}
    \item[\textbf{RQ1}] Does the proposed interface collect more feedback on exploratory items than the mix-in interface?
    \item[\textbf{RQ2}] Does the proposed interface provide better user experience than the mix-in interface?
    \item[\textbf{RQ3}] What are the differences between two interfaces in terms of how an exploratory item impacts perceived qualities?
\end{itemize}

\section{User Study Design} 
\label{sec_study}
In this section, we present our experiment design.
We first describe two interfaces used in the experiment--mix-in and revealed interfaces and then explain the data set, participants, and procedure of the experiment. 
Lastly, we provide detailed information on user-centric objectives of recommender systems shown in the survey questionnaire.

\subsection{Designing Recommender Interfaces} 
\label{sec_interface}
To answer our research questions, we designed two session-based interactive recommender interfaces--\textit{mix-in interface} and \textit{revealed interface}.
The mix-in interface was designed to allow the conventional recommendation navigation strategy, mixing exploratory items into a recommendation list. 
By hiding exploratory items from users, this interface can result in low transparency~\cite{Sinha02}. 
In contrast, the revealed interface presents exploratory and recommended items in two separate lists, informing users which items are personalized or not. 

\begin{figure*}
  \centering
  \includegraphics[width=0.95\textwidth]{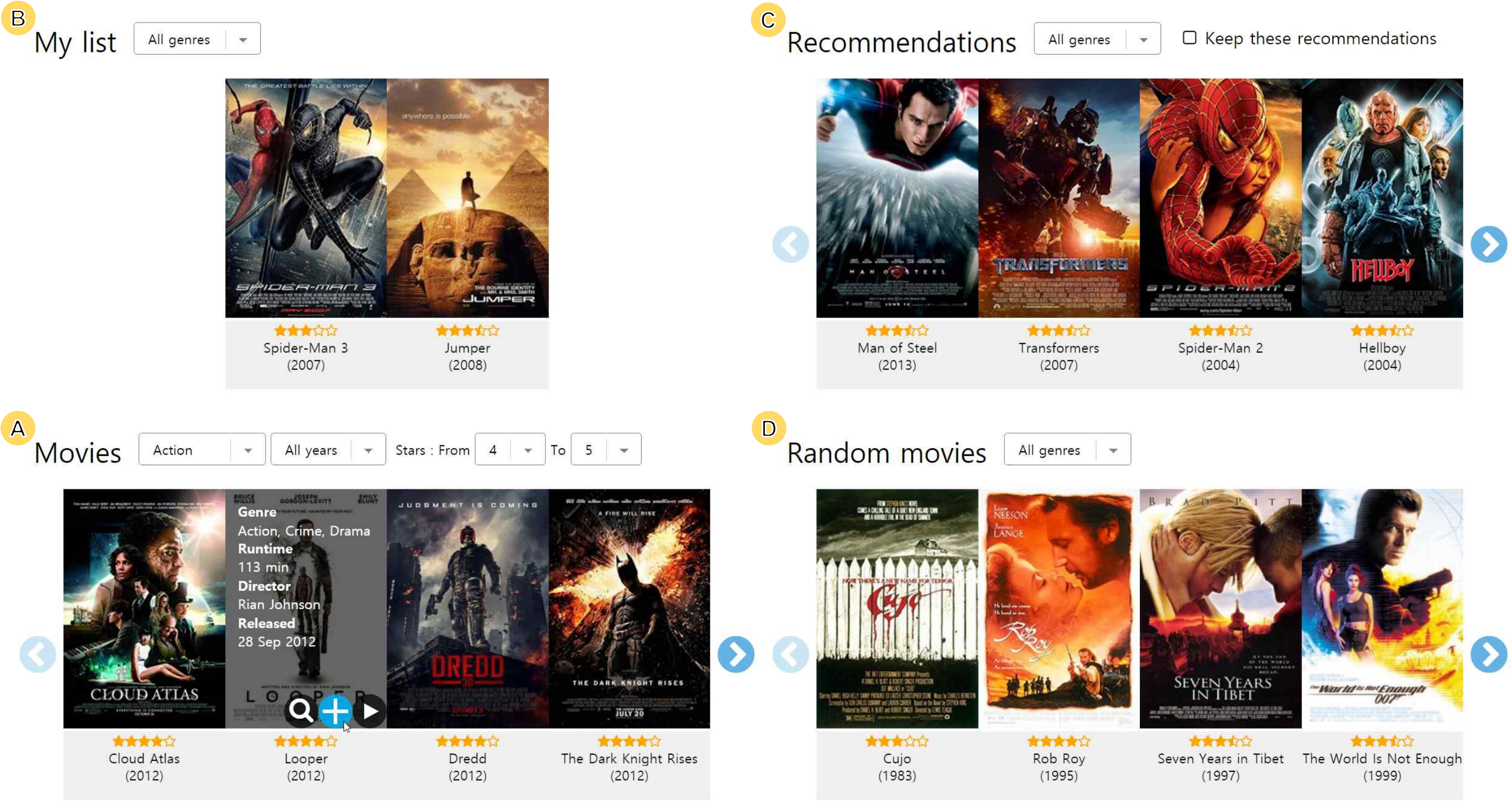}
  \caption{The revealed interface. Users can browse movies in (A), check candidate movies in (B), observe recommended movies in (C), and browse random movies in (D).
  }
  \label{fig_revealed_interface}
\end{figure*}

\autoref{fig_revealed_interface} shows the revealed interface used in the experiment, where four views (My list, Recommendations, Movies, and Random movies) are presented. 
In this revealed interface, `Movies' panel (A) is the main list, presenting 1027 movies in the movie data set.
``My list'' panel (B) is a list, where users place candidate movies that they currently consider to choose~\cite{Schnabel16}).
``Recommendations'' panel (C) recommends 50 movies for user preference based on the the movies in ``My list''.
The movies in the ``Movies'' panel are ranked by popularity for familiarity.
We considered a movie rated with more than three stars are popular in the MovieLens Latest dataset~\cite{Harper16}.
``Recommendations'' panel contains the 50 most similar movies for each movie in ``My list''.
We used cosine similarities between tag vectors from the Tag Genome project (i.e., content-based similarity computation)~\cite{Vig12}. 
In our test, this content-based computation recorded better performance than that of collaborative filtering methods, as the data set is not sufficiently large for the collaborative filtering methods. 
In the cosine similarity computation, all tag vectors were projected into a 25-dimensional space via Singular Value Decomposition for computational efficiency.

\begin{figure*}
  \centering
  \includegraphics[width=0.95\textwidth]{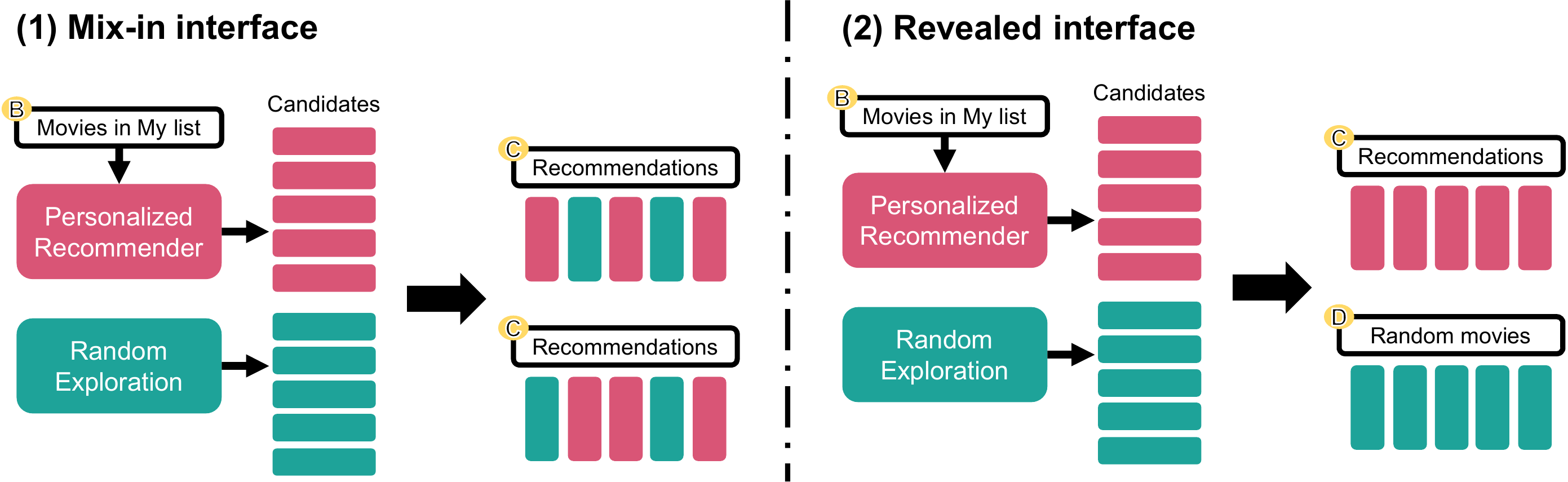}
  \caption{(1) The mix-in interface mixes randomly chosen exploratory items and personalized items into the recommendations lists in (\autoref{fig_revealed_interface} C), (2) The revealed interface presents exploratory and recommended items in two separated lists. The exploratory items are shown in the ``Random movies'' panel (\autoref{fig_revealed_interface} D).}
  \label{how_to_create_rec}
\end{figure*}

With the proposed interface, users can browse movies by clicking the arrow buttons (\includegraphics[width=0.020\linewidth]{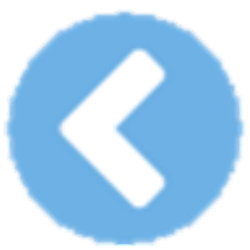}, \includegraphics[width=0.020\linewidth]{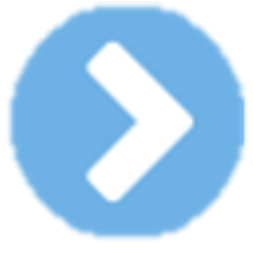}) in all panels and also filter the displayed items by the genre, year, and rating filters.
When a user hovers on a movie poster, detailed information of the movie is presented, such as the genre, run-time, director, and release date information with other buttons for displaying a short movie plot (\includegraphics[width=0.020\linewidth]{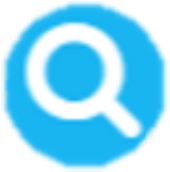}), adding the movie to the My list (\includegraphics[width=0.020\linewidth]{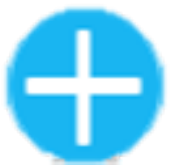}), and deciding the final movie (\includegraphics[width=0.020\linewidth]{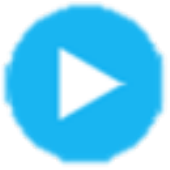}).
When a movie in the My list is hovered, a minus button (\includegraphics[width=0.020\linewidth]{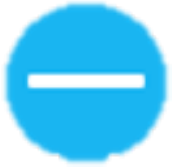}) is shown to remove the movie in the list, instead of the plus button. 
Whenever a movie is added or deleted in My list, the `Recommendations' panel (C) is automatically updated to reflect the addition or deletion.
To prevent the automatic update caused by user interactions on the recommendation list, users can select the check box (i.e., `Keep these recommendations') in (C).
Note that the mix-in interface is the same as the revealed interface, except that the ``Random movies'' panel (D) is replaced with another mix-in list and present a different movie set compared to the existing mix-in list. 
We filled 40\% of the recommendations with exploratory items, as done in Schnabel et al.'s study~\cite{Schnabel18}.
\autoref{how_to_create_rec} shows that two different presentation strategies of the mix-in and revealed interfaces.
Note also that exploratory items are randomly picked in the whole data collection.

\subsection{Movie Data}
We obtained movie data from ``MovieLens Latest Datasets''~\cite{Harper16} and collected detailed information on the obtained movies, such as poster images, synopsis, release dates show-times from OMDb API\footnote{http://omdbapi.com/}, an open movie database. 
Among the collected movies, we chose those that had 1800 or more ratings and were released after 1980 for ensuring familiarity. 
We also checked if movies' tag vectors were available in the Tag Genome dataset~\cite{Vig12} for computing relevance scores by cosine similarity. 
After the filtering process, we had 2054 movies in our inventory.
Then we split the movies in the inventory into two groups (1027 movies per groups) for each interface to prevent a user from having the same movie in the second session.
As an initial ranking method, we sort in a way that the participants see recent and highly-ranked movies by using year and rating information of the movies. 

\subsection{Survey Questions}
After each session finished (i.e., a user made the final choice), the participants filled in the questionnaire. 
All questions were answered on a 5-point Likert scale from strongly disagree (0) to strongly agree (4), except one free text question (``Feel free to leave any comment about the interface. (Optional)'').
This survey questions were adapted from the previous work in user-centric evaluation of a recommender~\cite{Knijnenburg12, Pu11} to the crowd sourcing setting~\cite{Kittur08}.
In our survey, we chose questions focusing on the following six properties of the recommendations: novelty, diversity, transparency, trust, satisfaction and accuracy.
In the following subsections, we introduce the variables measured in the questionnaire.

\textbf{Perceived Accuracy: }
Perceived accuracy is the degree to which users feel the recommendations match their interests and preferences~\cite{Pu11}. 
It is an overall assessment of how well the recommender has understood the users’ preferences and tastes.
We asked them for perceived accuracy in Question 1 (``The movies in the ``Recommendations'' panel were matched to my interests'') and Q2 (``The movies in the ``Recommendations'' panel were similar to those in the ``My list'' panel'').

\textbf{Novelty and Diversity: }
Novelty is extent to which users receive new and previously unseen recommendations~\cite{Pu11} and diversity means how different recommended items are from each other~\cite{Pu11}.
Often they are assumed to be related, because when a recommendation set is diverse, each item is novel to the rest of the recommendations in the set.
Conversely, a recommender system presenting novel results tends to result in global diversity over time in user experience. 
Moreover, a recommender system promote novel results tends to generate global diversity over time in the user experience and also increase the ''diversity of sales``~\cite{Fleder07} from the system perspective, in other words, eliminate the ``filter bubble''~\cite{Nguyen14}.
We measured novelty by Q3 (``The movies presented to me were novel.''). 
We also asked Q4 (``The movie I finally selected is different from my usual taste.'') to view impact of novelty on the users' final choice. 
We measured by asking Q5 (``The movies in the ``Recommendations'' panel were different to each other.'').

\textbf{Transparency and Trust: }
Other important properties of a recommender system are transparency and trust.
Transparency is why a specific recommendation was made and presented to the user, and trust indicates whether or not users find the whole system trustworthy~\cite{Pu11}.
We asked participants for transparency in two directions.
We asked them for transparency in the traditional way in question 11 (``I understood why the movies in the ``Recommendations'' panel were recommended.''), whereas we asked for transparency in the proactive setting in question 9 (``I feel in control of telling this movie recommender system what I want.'') and 10 (``I was able to steer the movie recommendations into the right direction.'').
We also asked for trust in questions 12 (``I understood why the movies were recommended to me.'') and 13 (``This recommender system can be trusted.'').

\textbf{Satisfaction: }
We asked participants for satisfaction in the following three directions: overall satisfaction, helpfulness and choice satisfaction.
Evaluating overall satisfaction determines what users think and feel while using a recommender system~\cite{Pu11}.
We asked them for overall satisfaction in question 14 (``Overall, I am satisfied with this system.''). 
A good recommender system should ultimately help the user with his/her decision, so we measured helpfulness of our recommender interfaces.
We asked them for helpfulness in question 15 (``This movie recommender system helped me find the movie I'd like to watch.'').
The goal of the task in our user study was picking a movie that participants would like to watch.
Therefore, a natural question to ask is how a presentation strategy for exploration (revealing or mix-in) affect participants' final choice and satisfaction of the choice.
We asked them for choice satisfaction in question 16 (`` I would watch the movie, when I have time. [with poster image of the chosen movie]'').

\subsection{Participants, Task, and Procedure of Experiment}
We recruited 117 participants from Amazon Mechanical Turk (MTurk) who were in U.S and had an approval rate higher than 95\%. 
The approval rate means that 95\% or more of a participant's previous submissions were approved by requesters. 
We prevented duplicated participation by asking them to submit their MTurk ID at the beginning of our experiment. 
Before the experiment, we required participants to use a minimum screen resolution of 1280 X 1024 pixel, a compatible web browser, and non-mobile devices (not tablet and smartphone) to control interfaces' layout and the number of items presented in the lists. 
Any participants who did not meet the requirements were not allowed to participate. 
Note that the compatible web browser refers one of the following list: Google Chrome ($\geq 4.0$), Microsoft Edge (Any version), Mozilla Firefox ($\geq 3.6$), Opera ($\geq 11.00$), and Safari ($\geq 4.0$).
The numbers in parentheses indicate the minimum version.

\begin{figure*}
  \centering
  \includegraphics[width=0.95\textwidth]{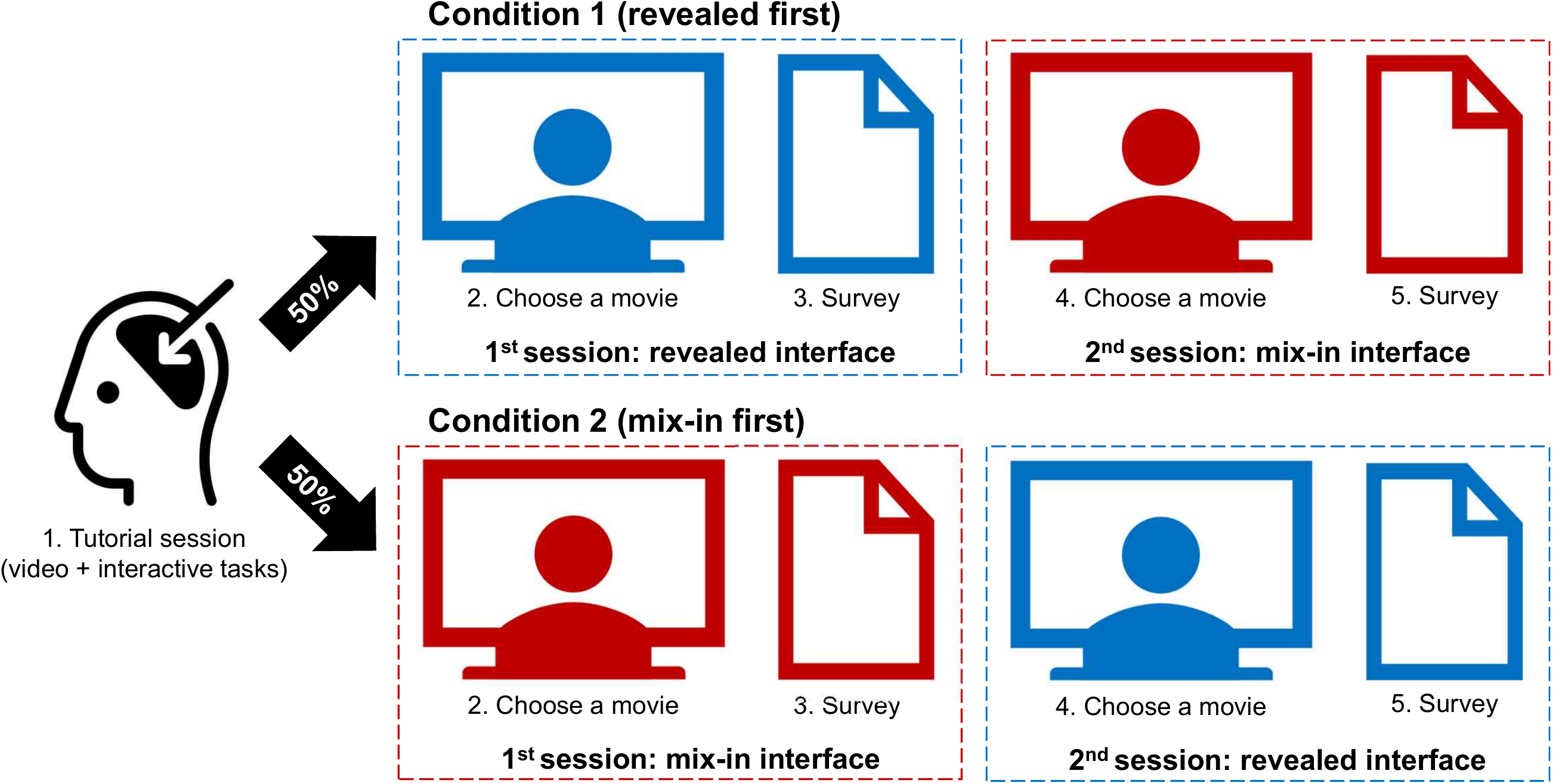}
  \caption{Participants are assigned one of two conditions with a 50\% probability. The difference between the two conditions is which interface is firstly shown to them. In each block, first, they take tutorial session composed of tutorial video summarizing the main functionality for the interface and interactive tutorial tasks. Second, they choose a movie which they want to watch in now in the interface. Last, they answer a questionnaire survey about the interface.}
  \label{fig:conditions}
\end{figure*}

\autoref{fig:conditions} describes our within-subject study, where participants performed the task twice with the different interfaces. 
To handle carry-over effects, we counterbalanced the conditions by assigning participants to one of the two conditions with 50\% probability. 
After participants agreed terms of the study, a tutorial session began, where we asked participants to watch a video and perform interaction tasks with the interfaces.
For example, the participants were asked to perform the following: ``Please choose your favorite genre in the Movies panel and add a movie into My list."
There was no time limit in the session and the participants were asked to complete 12 interaction tasks in the tutorial session to proceed. 

After the tutorial session finished and the movie data was loaded into their browser, the experiment began with the task--\textit{Please choose a movie you would like to watch \textbf{now}.} 
This task is based on the \textit{Find Good Items} task~\cite{Herlocker04}, or \textit{One-Choice} task~\cite{Schnabel16,Schnabel18,Schnabel19} which is a common task for evaluating recommender systems.
We recorded user interactions during the task for analysis. 
After each session finished, the participants filled out a survey questionnaire based on the five Likert scale (0-strongly disagree and 4-strongly agree), where questions on user experience and preference, and recommender interface quality were presented.
We utilized the previous questions proposed for user-centric evaluation of recommenders~\cite{Pu11}.

The 117 participants completed the experiment in about 12 minutes including a tutorial session, and we paid \$2.60, which is an effective wage of \$13.00 per hour and well above the US Federal minimum wage. 
The participants were 35.5 years old on average (SD=9.7, men: 73).
46\% of the participants reported that they have watched a movie more than once per week, followed by 35\% of people watching movies less often than that, but at least once a month. 
14\% of people said they watched a movie on a daily basis, and the remaining 5\% of the participants said they would watch a movie less than once a month.

In accordance with current best-practice guidelines for quality management of crowd-sourced data, we used a mix of outlier removal methods~\cite{Kittur08, Komarov13}. 
More specifically, we regarded the users who met the following criteria as outliers and removed their responses in our analysis:
\begin{itemize}
\item Users who stayed in each interface less than 15 seconds;
\item Users who were inactive for more than 90 seconds in a interface;
\item Users who reloaded the interface page;
\item Users who rated their final chosen item at 2 points or lower (Question 16 in our questionnaire).
\end{itemize}
In the end, we had 94 participants left for our result analysis.

\section{Experiment Results} \label{sec_result}
In this section, we present the experiment results.
First we investigate if our new interface allows us to collect more user feedback on exploratory items, compared to the conventional mix-in interface. 
Then we review effectiveness of the interface in terms of user-centric evaluation metrics. 
Lastly, we perform an in-depth analysis on interactions among the perceived qualities based on the user-centric evaluation metrics. 
\subsection{Revealed Interface Collected More Feedback on Exploratory Items (RQ1)}
The essential role of exploration is gathering information about a user’s tastes for future recommendations.
To evaluate if the revealed interface collected more user feedback on exploratory items than conventional approach, we compare the number of interactions on three interactions--\textit{Play}, \textit{Plus}, and \textit{Plot}.
\textit{Play} (\includegraphics[width=0.020\linewidth]{images/play.pdf}) interaction is the final choice with the interfaces, so the participants could have clicked it once.
If the count of this interaction is 0, it means, the participant chose a movie from the recommended movies, not from exploratory items. 
\textit{Plus} (\includegraphics[width=0.020\linewidth]{images/plus.pdf}) interaction adds the movies in the lists into the `My list'~\cite{Schnabel16}.
Thus the minimum count of \textit{Plus} is 0 and there is no limit in the maximum count. 
\textit{Plot} (\includegraphics[width=0.020\linewidth]{images/mag.pdf}) opens a pop-up window, showing a short movie plot and has the same maximum minimum count constraints as Plus.

\begin{table}[t]
\begin{center}
\begin{tabular}{@{}llll@{}}
\toprule
\textbf{Item type} & \textbf{Revealed} &\textbf{Mix-in} &  \textbf{\textit{p-value}}                 \\ \midrule
\textbf{Exploration}   & \textbf{1.71} ($\sigma$=2.10) & 0.97 ($\sigma$=1.41) & \textbf{0.0050**}    \\
\textbf{Recommendation}  & \textbf{3.88} ($\sigma$=1.60) & 3.78 ($\sigma$=1.55) & 0.6455             \\
\textbf{Both}          & \textbf{5.60} ($\sigma$=2.41) & 4.74 ($\sigma$=2.02) & \textbf{0.0098**}    \\ \bottomrule
\end{tabular}
\caption{Average interaction numbers on exploratory and recommendation items in each interface. We count three types of interactions-- \textit{Chosen}, \textit{Added} and \textit{Examined}. The participants left more interactions on exploratory items with the revealed interface.}
\label{table_log_agg}
\end{center}
\end{table}

\autoref{table_log_agg} shows the average interaction numbers on the three interaction types, where we find that the participants left different amounts of feedback based on interface types. 
Specifically, in case of the exploratory items, we find that the participants more interacted (t(94)=2.84, p<0.01) with revealed interface ($\mu$=1.71; $\sigma$=2.10), compared to mix-in interface ($\mu$=0.97; $\sigma$=1.41). 
However we do not find any difference in the number of interactions on the recommended items with both interfaces. 
When we compute the total number of interactions, we observe that the participants significantly interacted ($\mu$=5.60; $\sigma$=2.41) with the revealed interface (t(94)=2.61, p<0.01), compared to the mix-in interface ($\mu$=4.74; $\sigma$=2.02).

\subsection{Revealed Interface Provided Better User Experience (RQ2)}
\begin{table*}[t]
\centering
\resizebox{\textwidth}{!}{
\begin{tabular}{@{}lccc@{}}
\toprule
\multirow{2}{*}{\textbf{Questions}}                                          & \multicolumn{2}{c}{\textbf{Conditions}} & \multirow{2}{*}{\textit{\textbf{p}}} \\ \cmidrule(lr){2-3}
                                                                            & \textbf{revealed}   & \textbf{mix-in}   &                                      \\ \midrule
\textbf{Interface Adequacy}                                                 &                     &                   &                                      \\
1. ``The layout of this interface is adequate.''                                 & 3.351                 & 3.202               & 0.167                                 \\
2. ``This interface is comfortable for choosing movies.''                                & 3.511                & 3.543              & 0.802                                 \\
3. ``I became familiar with the movie recommender system very quickly.''                    & 3.394                 & 3.362              & 0.714                                 \\ \midrule
\textbf{Accuracy}                                                           &                     &                   &                                      \\
4. ``The movies in the `Recommendations' panel were matched to my interests.''                             & 3.064                 & 2.862              & 0.082                                 \\
5. ``The movies in the `Recommendations' panel were similar to those in the ``My list'' panel.''  & 2.872                 & 2.787              & 0.403                                 \\ \midrule
\textbf{Novelty}                                                            &                     &                   &                                      \\
6. ``The movies presented to me were novel.''                                   & 2.660                 & 2.255              & \textbf{0.001***}                                 \\
7. ``The movie I finally selected is different from my usual taste.''            & 1.617                 & 1.394              & 0.176                                 \\ \midrule
\textbf{Diversity}                                                          &                     &                   &                                      \\
8. ``The movies presented to me were different to each other.''                  & 2.957                 & 2.670              & \textbf{0.017*}                                 \\ \midrule
\textbf{Transparency}                                           &                     &                   &                                      \\
9. ``I feel in control of telling this movie recommender system what I want.''              & 3.309                 & 3.053              & \textbf{0.009**}                                 \\ 
10. ``I was able to steer the movie recommendations into the right direction.''          & 3.213                 & 2.989               & \textbf{0.026**}                                 \\
11. ``I understood why the movies in the `Recommendations' panel were recommended.''         & 3.309                 & 3.085               & \textbf{0.030*}                       \\ \midrule
\textbf{Trust}  &                     &                   &                                      \\
12. ``I am convinced of the items recommended to me.''                     & 2.968                 & 2.745              & \textbf{0.047*}                                 \\ 
13. ``This recommender system can be trusted.''                                & 3.170                 & 2.936              & \textbf{0.021*}                       \\ \midrule
\textbf{Satisfaction}  &                     &                   &                                      \\
14. ``Overall, I am satisfied with this system.''                             & 3.298                   & 3.011              & \textbf{0.007**}                                 \\
15. ``This movie recommender system helped me find the movie I’d like to watch.''                                 & 3.287                & 3.064              & \textbf{0.037*} \\                                 
16. ``I would watch the movie, when I have time. {[}with poster image of the chosen movie{]}''        & 3.649                & 3.596              & 0.455\\ \bottomrule
\end{tabular}
}
\caption{Aggregated answers for all survey questions. Answers were on a Likert scale from 0 (strongly disagree) to 4 (strongly agree). The last column has the \textit{p}-values for the independent sample t-test. (* \textit{p}<0.05, ** \textit{p}<0.01, *** \textit{p}<0.001)}
\label{survey-table}
\end{table*}

In this section we investigate how the interface affected perceived quality of the recommender systems based on the survey responses.
For this we run statistical tests on the survey responses and review user comments to find reasons of the results. 
For statistical analysis, we first test the assumption of equality of variances for user responses of each question through Bartlett's test~\cite{Snedecor89} and observe that all questions satisfy the equality of variances. 
Thus we run the sample t-test, whose results are presented in the last column of \autoref{survey-table}.

At first we see from the result table that there is no significant difference in the interface adequacy (Q1: t(94)=1.39, d=0.20, p>0.05; Q2: t(94)=0.25, d=0.04, p>0.05; Q3: t(94)=0.37, d=0.05, p>0.05).
This indicates that the participants did not have any issue or bias with the interface designs (e.g., layout, interaction methods) and felt comfortable during the experiment. 
We find several participants left positive comments on interface adequacy.  
\begin{quote} 
\textit{P12:} \textit{``This was a very intuitive and well-designed interface.''}

\textit{P2:} \textit{``This was very easy to use. I liked how the recommendations updated based of my selections in the movie panel.''}

\textit{P71:} \textit{``It seemed as though it worked pretty smoothly. It was easy enough to navigate and find a movie I wanted to watch.''}

\textit{P44:} \textit{``I liked being able to see the synopsis easily. Also, this interface is not heavily weighed down by code, very light and fast.''}
\end{quote}

Next we notice from the table that there is no statistically significant difference in the accuracy between the interfaces.
This indicates that both interfaces effectively deceived the users with their mix-in lists based on user preference.
This also confirms the result of Schnabel et al.'s study~\cite{Schnabel18} that users tend not to doubt exploration items in the list, when 40\% or less of the recommendations are exploratory items. 
We see two users thought the interfaces are very similar to each other in terms of recommendation quality. 
\begin{quote}
\textit{P93:} \textit{``This \emph{(mix-in interface)} felt close to the same as the original test \emph{(revealed interface)}, however there wasn't a random option - it was all just recommendations.'' \emph{(in the survey of mix-in interface)}}

\textit{P27:} \textit{``Algorithm seemed to be good, movies were similar to each other.'' \emph{(in the survey of the mix-in interface)}}
\end{quote}

For other questions, we see from \autoref{survey-table} that the participants expressed that the items in the revealed interface are more novel that those in the mix-in interface (Q6: t(94)=3.33, d=0.49, p<0.001). 
They also reported that the recommendation lists in the revealed interface are more diverse than those in the mix-in interface (Q8: t(94)=1.36, d=0.20, p<0.05).
But we do not see a significant difference, which implies that more exploratory items improve systems' novelty, but is not directly related to users' final choice, as Schnabel et al. also observe~\cite{Schnabel18}.
Reviewing user comments, we find the random movie panel contributes to enhancing both novelty and diversity of the recommender system. 
The participants identified the control for finding novel movies that are outside of their usual tastes as a strength of \textit{Random movies} panel.
\begin{quote} 
\textit{P28:} \textit{``I liked this \emph{(revealed interface)} one better. Sometimes I like to watch a movie outside of my usual genre and I could quickly browse the random movies and find one (with the interface). I have to go find the movie I picked now and watch it!''}

\textit{P72:} \textit{``Random movies panel is a good idea, sometimes I do not know what kind of movie I am in the mood for and can let the site choose for me.''}

\textit{P11:} \textit{``I choose the movie out of my usual tastes from the random movie list.''}
\end{quote}

we observe that the revealed interface received significantly higher scores for transparency (Q9: t(94)=2.19, d=0.32, p<0.05, Q10: t(94)=2.24; d=0.33; p<0.01, Q11: t=2.24; d=0.33; p<0.05) and trust (Q12: t(94)=2.00, d=0.29, p<0.05, Q13: t(94)=2.24, d=0.33, p<0.05) than the mix-in interface.
We think this result is obvious, because users can understand why the items are recommended to them in the revealed interface. 

Lastly, we find users were significantly more satisfied with the revealed interface compared to the mix-in interface. (Q14: t(94)=2.70, d=0.39, p<0.01).
We also notice that users thought that the revealed interface is more helpful during their item navigation (Q15: t(94)=2.1, d=0.31, p<0.05).
We think this happened, because users can notice random items in the recommendations and could not understand why the system recommended the movies, mismatched with movies in the `My list.' 
We find several complaints of those who gave low satisfaction scores to the mix-in interface as follows:

\begin{quote} 
\textit{P80:} \textit{``This interface \emph{(mix-in interface)} was somewhat strange, since the recommendations looked like random.''}

\textit{P66:} \textit{``Algorithm seemed to be bad, the recommended movies were very dissimilar to each other.''}

\textit{P49:} \textit{``The recommendations in this interface \emph{(mix-in interface)} did not match the genre that I chose and they were very random movie genres'' \emph{(in the survey of the mix-in interface)}}

\textit{P55:} \textit{``The interface \emph{(mix-in interface)} worked fine, but the recommendations seem as random.'' \emph{(in the survey of the mix-in interface)}}
\end{quote}

For Q16 that asked satisfaction on the final choice and we do not see any significant difference. 
Seeing the score distribution, we find that users gave similar high scores (3.649 and 3.596 for the revealed and mix-in interfaces). 
This implies that they were satisfied with their final choice, regardless of the interface used for deciding their final choice.
This result is consistent with the finding from of Schnabel et al.'s study~\cite{Schnabel18}.

To sum up, we answer RQ2 in this section and observe that the revealed interface provides better user experience with respect to novelty, diversity, transparency, trust and satisfaction.
But we still have question on the relationships among the objectives and which objectives are directly related to improving user satisfaction (Q3). 
To further investigate the question, we perform correlation and structural modeling in the following sections. 

\subsection{Revealed Interface Positively Affects Perceived Qualities (RQ3)}
In this work, we assume that our interface can affect three objectives--transparency, diversity, and novelty due to innate characteristic of the proposed interface. 
To investigate the impact of the interface on the three objectives and causal relationships among the objectives, we perform multi-group path analysis based on structure equation modeling (SEM)~\cite{Kaplan08}. 
To conduct the analysis, we first need to check if conditions for the anlaysis are satisfied with our experiment data.
In the next sections, we present descriptive statistics to exam outliers, missing values, and normality of the variables. 
Then we confirm if all of the conditions are met by performing several analyses, including reliability analysis, correlation analysis, and factor analysis on the variables.
Lastly, we report fitness indices of our structural equation model and present our multi-path analysis result on the overall impact of using the proposed interface on the perceived quality of recommender systems. 
We use R (Version: 3.6.1, released in 2019-07-05) and Lavaan (Version: 0.6.4) for computation. 

\begin{table}[t]
\resizebox{\textwidth}{!}{
\begin{tabular}{@{}ccccccccccccccccc@{}}
\multicolumn{1}{l}{}          & \multicolumn{1}{l}{} &  & \multicolumn{4}{c}{\textbf{Total} \textit{(N=188)}} &  & \multicolumn{4}{c}{\textbf{Revealed} \textit{(N=94)}} &  & \multicolumn{4}{c}{\textbf{Mix-in} \textit{(N=94)}} \\ \cmidrule(lr){4-7} \cmidrule(lr){9-12} \cmidrule(l){14-17} 
\textbf{Latent variables}              & \textbf{Observed variables}   &  & \textbf{Mean} & \textbf{SD}   & \textbf{Skewness} & \textbf{Kurtosis} &  & \textbf{Mean}  & \textbf{SD}    & \textbf{Skewness} & \textbf{Kurtosis} &  & \textbf{Mean} & \textbf{SD}   & \textbf{Skewness} & \textbf{Kurtosis} \\ \cmidrule(r){1-2} \cmidrule(lr){4-7} \cmidrule(lr){9-12} \cmidrule(l){14-17} 
\multirow{3}{*}{Exploration}  & Play                 &  & 0.19 & 0.40 & 1.84     & 2.10     &  & 0.27  & 0.47  & 1.35     & 0.57     &  & 0.11 & 0.31 & 2.51    & 4.36     \\
                              & Plus                 &  & 0.94 & 1.42 & 1.44     & 0.92     &  & 1.15  & 1.59  & 1.12     & -0.15    &  & 0.72 & 1.20 & 1.78    & 2.56    \\
                              & Plot                 &  & 0.22 & 0.52 & 2.32     & 4.40     &  & 0.30  & 0.58  & 1.78     & 2.03     &  & 0.14 & 0.43 & 3.16    & 9.42     \\
\cmidrule(r){1-2}
\multirow{2}{*}{Accuracy}     & Question 4           &  & 2.96 & 0.80 & -1.20    & 2.13     &  & 3.06  & 0.72  & -0.96    & 1.68     &  & 2.86 & 0.86 & -1.23    & 1.80     \\
                              & Question 5           &  & 2.83 & 0.70 & -0.90    & 1.21     &  & 2.87  & 0.69  & -0.80    & 1.14     &  & 2.79 & 0.70 & -0.98    & 1.14     \\
\cmidrule(r){1-2}
\multirow{2}{*}{Novelty}      & Question 6           &  & 2.46 & 0.85 & -0.46    & 0.45     &  & 2.66  & 0.87  & -0.73    & 0.59     &  & 2.26 & 0.79 & -0.34    & 0.88     \\
                              & Question 7           &  & 1.51 & 1.13 & 0.58     & -0.47    &  & 1.62  & 1.28  & 0.43     & -1.13    &  & 1.39 & 0.95 & 0.63     & 0.52     \\
\cmidrule(r){1-2}
Diversity                     & Question 8           &  & 2.81 & 0.83 & -0.60    & -0.02    &  & 2.96  & 0.85  & -0.84    & 0.34     &  & 2.67 & 0.78 & -0.43    & -0.17    \\
\cmidrule(r){1-2}
\multirow{3}{*}{Transparency} & Question 9           &  & 3.18 & 0.67 & -0.54    & 0.48     &  & 3.31  & 0.62  & -0.58    & 0.66     &  & 3.05 & 0.69 & -0.45    & 0.28     \\
                              & Question 10          &  & 3.10 & 0.69 & -0.42    & 0.10     &  & 3.21  & 0.70  & -0.87    & 1.28     &  & 2.99 & 0.66 & 0.01     & -0.76    \\
                              & Question 11          &  & 3.20 & 0.71 & -0.84    & 1.13     &  & 3.31  & 0.67  & -0.86    & 1.19     &  & 3.09 & 0.73 & -0.79    & 0.99     \\
\cmidrule(r){1-2}
\multirow{2}{*}{Trust}        & Question 12          &  & 2.86 & 0.77 & -0.38    & -0.13    &  & 2.97  & 0.84  & -0.60    & -0.11    &  & 2.74 & 0.69 & -0.22    & -0.06    \\
                              & Question 13          &  & 3.05 & 0.70 & -0.45    & 0.21     &  & 3.17  & 0.70  & -0.61    & 0.47     &  & 2.94 & 0.68 & -0.32    & 0.12     \\
\cmidrule(r){1-2}
\multirow{2}{*}{Satisfaction} & Question 14          &  & 3.15 & 0.74 & -0.72    & 0.51     &  & 3.30  & 0.67  & -0.63    & 0.19     &  & 3.01 & 0.78 & -0.68    & 0.35     \\
                              & Question 15          &  & 3.18 & 0.74 & -0.69    & 0.34     &  & 3.29  & 0.60  & -0.49    & 0.93     &  & 3.06 & 0.84 & -0.55    & -0.43    \\
                              & Question 16          &  & 3.62 & 0.49 & -0.50    & -1.76    &  & 3.65  & 0.48  & -0.61    & -1.64    &  & 3.60 & 0.49 & -0.38    & -1.87      
\end{tabular}
}
\caption{The mean, standard deviation, skewness, and kurtosis of variables for each group.}
\label{desc-survey}
\end{table}

\subsubsection{Descriptive statistics for variables}
\autoref{desc-survey} shows the objectives chosen as variables for the analysis.
The variable data are from interaction logs and survey questions. 
Note that all variables from the questionnaire have values from zero to four, the \textit{Play} variable can be either zero or one, and the minimum value of \textit{Plus} and \textit{Plot} is zero, and the maximum can be up to infinity.
Next we perform a normality test before we proceed next analysis steps. 
\autoref{desc-survey} presents the mean, standard deviation, skewness and kurtosis of the variables in consideration of total, revealed, and mix-in interface users. 
From the statistics, we find that the variables closely follow a normal distribution with the mean and standard deviation close to zero and one, respectively. 
We also observe that all variables, except \textit{plot} pass a more rigorous normality test with the conditions of $|skewness|<3$ and  $|kurtosis|<8$~\cite{West95}. 
As \textit{plot} does not pass the normality test, we exclude it in the structural equation modeling. 

\begin{table}[t]
\resizebox{\textwidth}{!}{
\begin{tabular}{@{}ccccccccc@{}}
\toprule
& \textbf{Exploration} & \textbf{Accuracy} & \textbf{Novelty} & \textbf{Diversity} & \textbf{Transparency} & \textbf{Trust}   & \textbf{Satisfaction} & \textbf{Cronbach's \textit{α}} \\ \midrule
\textbf{Exploration}  & 1        & 0.11     & 0.46***  & 0.31***  & 0.32***  & 0.33***  & 0.10     & 0.81         \\
\textbf{Accuracy}     & 0.11     & 1        & 0.15*    & 0.41***  & 0.54***  & 0.57***  & 0.56***  & 0.78         \\
\textbf{Novelty}      & 0.46***  & 0.15*    & 1        & 0.33***  & 0.20***  & 0.41***  & 0.05     & 0.81         \\
\textbf{Diversity}    & 0.31***  & 0.41***  & 0.33***  & 1        & 0.52***  & 0.51***  & 0.28***  & 0.77         \\
\textbf{Transparency} & 0.32***  & 0.54***  & 0.20***  & 0.52***  & 1        & 0.71***  & 0.64***  & 0.75         \\
\textbf{Trust}        & 0.33***  & 0.57***  & 0.41***  & 0.51***  & 0.71***  & 1        & 0.65***  & 0.73         \\
\textbf{Satisfaction} & 0.10     & 0.56***  & 0.05     & 0.28***  & 0.64***  & 0.65***  & 1        & 0.78         \\ \bottomrule
\end{tabular}}
\caption{Correlation between latent variables for the complete sample \textit{(N=188)}. (* p<0.05, ** p<0.01, *** p<0.001)}
\label{corr_total}
\end{table}

\subsubsection{Correlation analysis}
Before conducting the structural equation modeling, we perform correlation analysis for providing an outline of the relationships among the variables.
The rationale behind this correlation analysis is that if the correlation between two variables is not significant, the result in the multi-group path analysis result is also meaningless.
\autoref{corr_total} shows the computed correlation results (N=188), where we find that many positive correlations between the variables. 
At first, we can find that the relationships between variables that we proposed in \autoref{proposed_path_model} have positive correlations. 
Specifically, exploration is positively correlated with all of the transparency (r=0.32, p$<$0.001), diversity (r=0.31, p$<$0.001), and novelty (r=0.46, $p$<0.001). 
This result means that exploratory items shown in the revealed interface contribute to improving user experience in transparency, diversity, and novelty perspectives. 
We can also confirm that the proposed causality model based on prior literature survey (\autoref{proposed_path_model}) is valid, as we find that all paths show moderate to strong correlations (Transparency->Diversity~\cite{Castagnos13}: r=0.52, p$<$0.001), (Transparency->Trust~\cite{Barraza17,Pu11,Schnabel18}: r=0.71, $p$<0.001),
(Transparency->Accuracy~\cite{Barraza17,Schnabel18}: r=0.54, $p$<0.001), 
(Trust->Satisfaction~\cite{Pu11}: r=0.65, p$<$0.001), 
(Accuracy->Satisfaction~\cite{Pu11}: r=0.56, $p$<0.001), 
(Diversity->Satisfaction~\cite{Pu11}: r=0.28, $p$<0.001). 
The relationship between the novelty and diversity has been controversial (\autoref{proposed_path_model} dotted line)~\cite{Castells11,Ekstrand14,Niu18}. 
Our correlation analysis result shows that they are positively correlated. 
(Novelty->Diversity~\cite{Pu11}: r=0.33, $p$<0.001). 
But we do not find any evidence of the relationship between novelty and user satisfaction (r=0.05, p$>$0.05).  

As we confirm the relationships among variables with the correlation analysis, we check multicollinearity of the variables, before proceeding to structural equation modeling.
Multicollinearity is a statistical phenomenon in which two or more predictor variables in multiple regression models are highly correlated and high multicollinearity (e.g., higher than 0.8~\cite{Maddala88,Chen19}) causes disturbance and unreliability in the regression analysis~\cite{Donald67}.
In our inspection, we see that all correlation values are below 0.8.
Next we run another multicollinearity diagnosis by computing variance inflation factor (VIF) and find all variables have low VIF values (\textit{Accuracy}: 1.65, \textit{Novelty}: 1.32, \textit{Diversity}: 1.66, \textit{Transparency}: 1.33, \textit{Trust}: 4.67, \textit{Satisfaction}: 3.58).
Note that VIF is widely used to assess the degree of collinearity among independent variables and when VIF values are less than 10, we can think there is no multicollinearity issue in the analysis~\cite{Hair95}. 
Lastly we measure Cronbach's alpha that indicates the degree to which a set of items measures a single unidimentional latent construct and  each latent variable is measured consistently by the set of measurement variables belong to the latent variable.
We find from \autoref{corr_total} (last column) that Cronbach's \textit{α} values of all variables show acceptable ($\textit{α}>0.60$)~\cite{Moss98, Hair2006,Griethuijsen15} or good levels ($\textit{α}>0.71$)~\cite{Griethuijsen15,Tavakol11}.

\begin{table}[t]\centering \caption{Measurement invariance test result. (* p<0.05, ** p<0.01, *** p<0.001)}
\begin{threeparttable}
\begin{tabular}{@{}ccccccccc@{}}
\toprule\midrule
              & Df  & CFI    & RMSEA    & Chisq  & Chisq diff & Df diff & Pr(\textgreater{}{${\chi}^{2}$}) &  \\ \midrule
$M1$ & 136 & 0.910 & 0.094 & 249.23 &            &         &                         &  \\
$M2$ & 143 & 0.910 & 0.092 & 256.38 & 7.1538     & 7       & 0.413049                &  \\
$M3$ & 150 & 0.913 & 0.088 & 259.43 & 3.0460     & 7       & 0.880711                &  \\
$M4$ & 162 & 0.900 & 0.091 & 287.67 & 28.2483    & 12      & 0.005087**              &  \\
$M5$ & 169 & 0.891 & 0.093 & 307.00 & 19.3224    & 7       & 0.007235**              &  \\
\bottomrule\addlinespace[1ex]
\end{tabular}
\begin{tablenotes}\footnotesize
\item[M1] Unconstrained model (The same factor structure is imposed on all groups).
\item[M2] The factor loadings are constrained to be equal across groups.
\item[M3] The factor loadings and intercepts are constrained to be equal across groups.
\item[M4] The factor loadings, intercepts and residual variances are constrained to be equal across groups.
\item[M5] The factor loadings, intercepts, residual variances and means are constrained to be equal across groups.
\end{tablenotes}
\end{threeparttable}
\label{tbl-mi}
\end{table}

\subsubsection{Configural and measurement invariance tests}
We further conduct two more tests--configural and measurement invariance tests. 
We inspect configural invariance to see if the factor structure (i.e., the proposed path model) is the same across the two groups.
To guarantee the configural invariance, we use the goodness-of-fit indices--Tucker–Lewis index (TLI)~\cite{Tucker73}, comparative fit index (CFI)~\cite{Bentler90}, root mean square error of approximation (RMSEA)~\cite{Hu99}, and standardized root mean squared residual (SRMR)~\cite{Byrne13}.
After computation of the indices, we have $CFI=0.910$, $TLI=0.880$, $SRMR=0.072$, $RMSEA=0.094$, confirming this result satisfies the criteria proposed by Hong et al.~\cite{Hong03}, as follows: 
\begin{itemize}
\item The Tucker–Lewis index (TLI) ≥ 0.80~\cite{Tucker73};
\item The comparative fit index (CFI) ≥ 0.80~\cite{Bentler90};
\item The standardized root mean squared residual (SRMR) ≤ 0.08~\cite{Hu99};
\item The root mean square error of approximation (RMSEA) ≤ 0.10~\cite{Byrne13}.
\end{itemize}

Next we investigate the measurement invariance to confirm that the users in the two groups perceive the questions (i.e., constructs) in the same way.
\autoref{tbl-mi} provides information for investigating the measurement invariance with four levels, each of these levels builds upon the previous model by introducing additional equality constraints on model parameters to achieve stronger forms of invariance.  
In the investigation process, each set of new parameters is tested and the parameters known to be invariant from previous levels are constrained. 
Specifically, M1 is set as the proposed path model. 
Then we have M2 by constraining M1's factor loadings to be equal across the groups. 
Here the factor loadings are the constructs of the survey questions. 
In M3, factor loadings and intercepts additionally are constrained, compared to M2. 
In M4, residual variances are additionally constrained, compared to M3. 
M5 is the model, where all of the factor loadings, intercepts, residual variances and means are constrained.

In the measurement invariance test, the best case is called strict measure invariance, where all of the four levels' conditions are met.
But it is generally agreed that expecting equality in residual variances across groups is not realistic and the such strict level of invariance is rarely achieved in practice~\cite{Rens15}.
Instead, metric invariance is usually required for meaningful comparison between groups, which means M2 and M3 are same.  
For assessing the measure invariance, we conduct the chi-square test and find we achieve the metric invariance, because we do not see that there is no significant difference between not only M1 and M2, but also M2 and M3. 
But we see M3 and M4 are significantly different. 

To sum up, we review the relationships between the variables, as proposed in \autoref{proposed_path_model} with correlation anlaysis and check that our data does not have the multicollinearity issue. 
We additionally confirm that the two groups are not only satisfy the configural invariance, but also the measure invariance at the level of metric invariance.
These results satisfy the criteria for multi-group path analysis~\cite{Vandenberg00}, whose result is presented in the next section. 

\begin{figure}[t]
  \centering
  \includegraphics[width=0.95\textwidth]{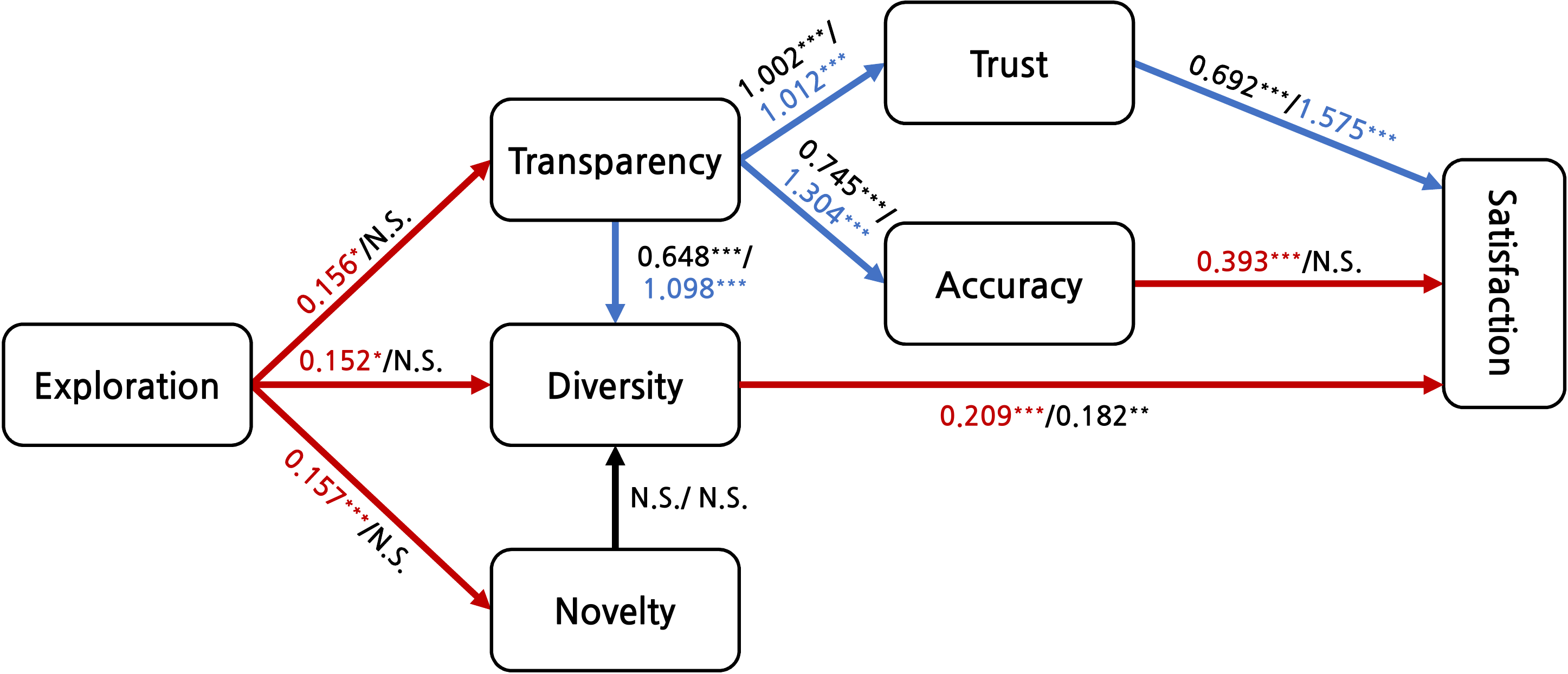}
  \caption{Multi-group path model analysis between the revealed and mix-in groups. Red highlights higher $\beta$ coefficients in the revealed group, and blue highlights higher $\beta$ coefficients in the mix-in group. A value to the left of the slash mark (`/') indicates a $\beta$ coefficient of the revealed group, and the right is for the mix-in group. N.S. indicates non-significant values ($p>0.05$). (* \textit{p}<0.05, ** \textit{p}<0.01, *** \textit{p}<0.001)}
  \label{fig:resulted-model}
\end{figure}

\subsubsection{Multi-group path analysis}

Multi-group path analysis allows us to explore causal relationships among variables and we find in the previous sections that all conditions are met for running the multi-group path analysis. 
Next we compare the structural equations models of the two interfaces ($N=94$). 
At first, we find that the two path models are significantly different from each other (${\chi}^{2}\big(68\big) = 49.97$, $p < 0.001$), which implies that the coefficients of the paths in the models are much different each other. 

\autoref{fig:resulted-model} presents a summary of the multi-path analysis. 
Here the arrows represent causal relationships between two objectives and the numbers on the arrows show the beta values of the revealed and mix-in interfaces, respectively in the analysis. 
We assign red on the arrows when the beta values of the revealed interface are higher than those of the mix-in interfaces. 
When the mix-in interface has a higher beta value, we present blue arrows.

We identify three major differences of the two path models. 
First, we see that the number of interactions on exploratory items significantly impacts three objectives--transparency($\beta=0.156, p<0.001$), diversity ($\beta=0.152, p<0.05$), and novelty ($\beta=0.157, p<0.001$), but this impact only exists in the revealed interface. 
In addition, we observe that accuracy ($\beta=0.393, p<0.001$) and diversity ($\beta=0.209, p<0.001$) are the two major factors, affecting satisfaction of recommender systems, when the revealed interface is used. 
In contrast, trust is the major factor with a large coefficient that directly impact satisfaction in the mix-in interface. 
We can also consider transparency as an influencing factor to satisfaction due to its connection to trust (coefficient: 1.002) in the mix-in interface. 
But we do not find any impact of novelty to diversity in both interfaces. 

\section{Analysis, Implications, Discussion of the Results} 
\label{sec_discussion}
In this section, we summarize findings and implications of the study.
First we report that the conventional concern--using exploration items lead to user dissatisfaction~\cite{Schnabel18} can be overcome by using a novel item presentation strategy.  
For example, in this study, we use an interface where users can notice which items are exploration items. 
The experiment results indicate that doing so lead to higher user satisfaction compared the conventional mix-in interface that hides the exploration items. 
We report that our approach works better than the mix-in interface in terms of improving user experience and collecting implicit user feedback on exploration items without impacting perceived accuracy of recommender systems.
In other words, both the revealed and mix-in interfaces are not significantly different, when evaluated with perceived accuracy only, but our revealed interface provides better user experience when considering other beyond-accuracy factors.

The multi-group path analysis results indicate that the number of user interaction on exploratory items positively affect transparency, diversity, and novelty objectives of recommender systems. 
This suggests that a recommender system that reveals exploratory makes users feel the recommendation list is novel, diverse, and transparent. 
We also find that transparency indirectly affects user satisfaction through trust, while diversity directly does. 
Eventually, exploratory items presented in the revealed interface contribute to improving user satisfaction, but they do not in the mix-in interface.

In \autoref{proposed_path_model}, we summarize relationships of objectives based on previous work, which says improving novelty  positively affects diversity~\cite{Pu11, Ekstrand14}. 
But in our multi-group path analysis, we do not see such effect between the objectives. 

There has been controversy in whether novelty has a positive~\cite{Ekstrand14} or negative~\cite{Chen19, Pu11} effect on other factors.
Some researches indicated a positive impact of novelty on user satisfaction~\cite{Pu11, Chen19}, but they contradict the findings of ~\cite{Ekstrand14}. 
As discuss in ~\cite{Kaminskas17}, this contradiction may be due to the studied product domain (e-commerce products vs. movies) or the formulation of novelty-related survey question (positive such as ``The item recommended to me is novel'' tone in  ~\cite{Pu11} vs. negative tone such as ``Which list has more movies you would not have thought to consider?'' in ~\cite{Ekstrand14}).
Thus, more studies should be conducted to verify these confounding effects.

To sum up, we show that algorithmic exploration can be used for two main goals--gathering a user's tastes and improving user experience such as novelty, diversity and satisfaction, and to achieve the goals, our result suggest to reveal the exploration process to users.

\section{Limitations and Future Work} \label{sec_limitations}
In addition, we found our revealed interface improve user experience in the terms of user-centric evaluation and collected more implicit feedback on exploratory items. 
The experiment of our work used a session-based recommender setting, so we observed this phenomenon in only the short-term session.
It would be interesting to connect our results to user behaviour in the long-term user study or the wild. 

Also, there is a open question, the strength of domain-dependent effects. 
Since picking a movie is a quite visual task, there is more research needed to study other domains, for example, the task of having to pick a music or book where less information can be inferred from the pictorial representation of an item.

There is the other open questions, how much interactivity affects how people perceive recommended and exploratory items.
There are small yet growing researches studying user behaviour and perception in an interactive recommender~\cite{Schnabel16,Schnabel18,Schnabel19}, where follow many findings from traditional long-term user studies.
However, more studies are needed to measure and model the precise connections between traditional static recommendations and interactive recommendations. 
As an intermediate work, combining interactive recommendation with approaches for session-based recommendation~\cite{Hidasi16,Wu16b,Jannach15} would be interesting future work.
\section{Conclusions}
\label{sec_conclusions}
In this work, we firstly measured the effect of hiding and revealing exploration process from/to users, and compared them in terms of user experience and the quantity of implicit feedback.
Our results consistently show revealing exploration process to users (\textit{revealed exploration}) is a better strategy than hiding it from them (\textit{mix-in exploration}). 
Specifically, revealed exploration not only got higher scores in novelty, diversity, transparency, trust and satisfaction in a user-centric evaluation, but also gathered more implicit feedback on the exploratory items, which is original purpose of exploration.
Our findings provide the following practical advice when designing a interface that utilizes learning algorithm based on users' feedback signals: \textit{``Do not hide exploration from users, instead reveal it''}. 
We hope that our work will facilitate closer collaboration between machine learning specialists and UI/UX designers.

\bibliographystyle{acm}
\bibliography{references} 

\end{document}